\documentclass{aipproc}
\layoutstyle{6x9}
\usepackage{graphics}

\def\Ref#1{\eqref{#1}}
\def\Ps#1#2{\centerline{\scalebox{#1}{\includegraphics{#2}}}}
\def\PS#1#2#3{\centerline{
  \scalebox{#1}{\includegraphics{#2}}~~~
  \scalebox{#1}{\includegraphics{#3}}
}}

\def\BA{\begin{eqnarray}}
\def\BE{\begin{equation}}
\def\BF{\begin{figure}[htb]}
\def\BT{\begin{table}[htb]}
\def\EA{\end{eqnarray}}
\def\EE{\end{equation}}
\def\EF{\end{figure}}
\def\ET{\end{table}}
\def\vr{{\vec r\,}} \def\rT{r_T} \def\vrT{{\vec r_T}}
\def\vp{{\vec p\,}} \def\pT{p_T} \def\vpT{{\vec p_T}}
  
\def\la{\langle}
\def\ra{\rangle}

\def\mb{\,\mbox{mb}}

\def\fm{\,\mbox{fm}}
\def\GeV{\,\mbox{GeV}}

\def\eps{\varepsilon}

\def\Jpsi{{J\!/\!\psi}}
\def\sqq{{\sigma_{q\bar q}}}
\def\aem{{\alpha_{\mbox{\tiny em}}}}

\begin{document}

\title{Electroproduction of Charmonia\\off Protons and Nuclei}
\author{Yu.P. Ivanov}{
     address={Institut f\"ur Theoretische Physik der Universit\"at, 69120 Heidelberg, Germany},
  altaddress={Joint Institute for Nuclear Research, Dubna, 141980 Moscow Region, Russia}
}
\author{B.Z. Kopeliovich}{
     address={Institut f\"ur Theoretische Physik der Universit\"at, 93040 Regensburg, Germany},
  altaddress={Max-Planck Institut f\"ur Kernphysik, Postfach 103980, 69029 Heidelberg, Germany}
}
\author{A.V. Tarasov}{
     address={Joint Institute for Nuclear Research, Dubna, 141980 Moscow Region, Russia}
}
\author{J. H\"ufner}{
     address={Institut f\"ur Theoretische Physik der Universit\"at, 69120 Heidelberg, Germany}
}
\begin{abstract}
Elastic virtual photoproduction of charmonia on nucleons is calculated in
a parameter free way with the light-cone dipole formalism and the same
input: factorization in impact parameters, light-cone wave functions
for the photons and the charmonia, and the universal phenomenological
dipole cross section which is fitted to other data. The charmonium wave
functions are calculated with four known realistic potentials, and two
models for the dipole cross section are tested. Very good agreement
with data for the cross section of charmonium electroproduction is found
in a wide range of $s$ and $Q^2$. Using the ingredients from those
calculations we calculate also exclusive electroproduction of charmonia
off nuclei. Here new effects become important, (i) color filtering of
the $c\bar c$ pair on its trajectory through nuclear matter, (ii)
dependence on the finite lifetime of the $c\bar c$ fluctuation (coherence
length) and (iii) gluon shadowing in a nucleus compared to the one in a
nucleon. Total coherent and incoherent cross sections for C, Cu and Pb as
functions of $s$ are presented. The results can be tested with future 
electron-nucleus colliders or in the peripheral collisions of
relativistic heavy ions.
\end{abstract}

\maketitle


\section{Introduction}

In contrast to hadro-production of charmonia, where the mechanism is
still debated, electro(photo)production of charmonia seems better
understood: the $c\bar c$ fluctuation of the incoming real or virtual
photon interacts with the target (proton or nucleus) via the dipole
cross section $\sqq$ and the result is projected on the wave function
of the observed hadron \cite{KZ}. The aim of this paper is not to
propose a conceptually new scheme, but to calculate within this
approach as accurately as possible and without any free parameters.
Wherever there is room for arbitrariness, like form for the color
dipole cross section and for for charmonium wave function, we use
and compare other author's proposals, which have been tested on 
different data.

In the light-cone (LC) dipole approach the virtual photoproduction
of charmonia (here $\Psi$ stands for $\Jpsi$ or $\psi'$) looks as
shown in Fig.~\ref{Fig-gp} \cite{KZ}.
\BF
\Ps{0.45}{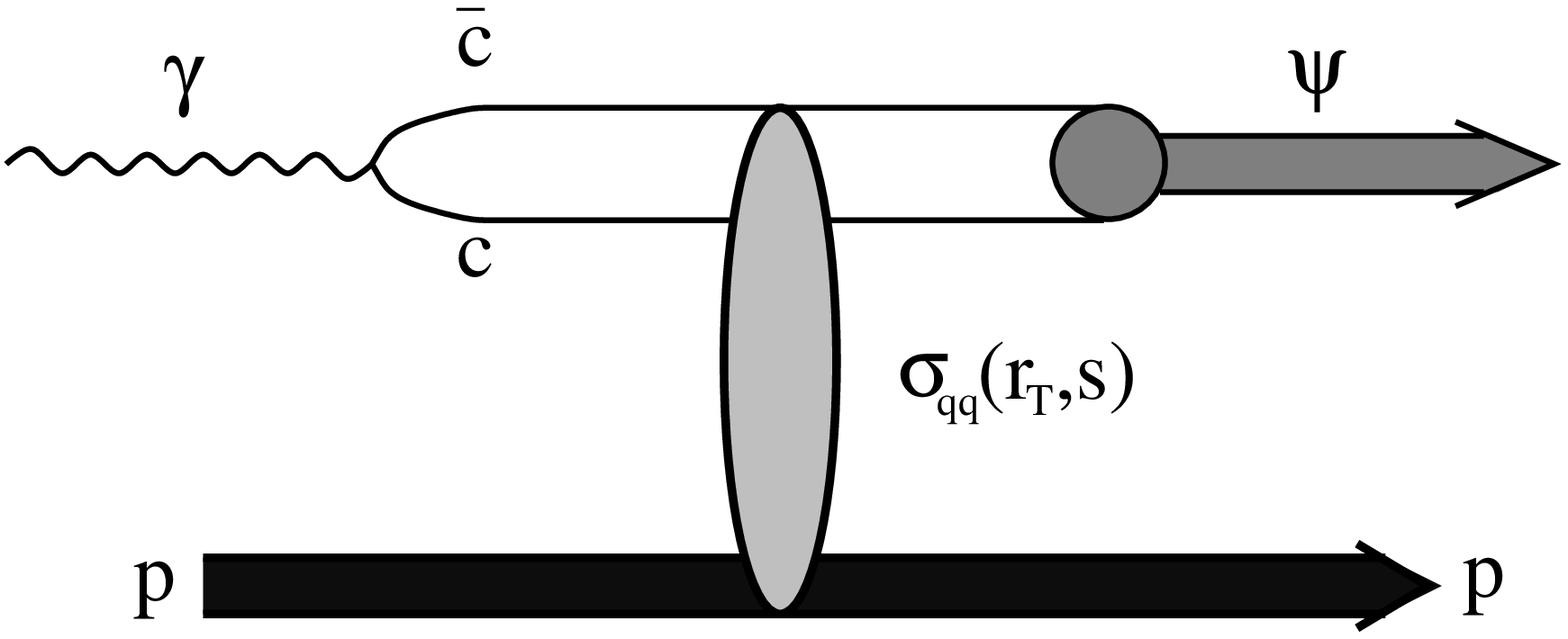}
\caption{
  \label{Fig-gp}
  Schematic representation of the amplitude for the reaction
  $\gamma p\to\Psi p$ in the rest frame of the proton. The
  $c\bar c$ fluctuation of the photon with transverse separation
  $\rT$ and c.m. energy $\sqrt s$ interacts with the target proton
  via the dipole cross section $\sqq(\rT,s)$ and produces a $\Psi$.
}
\EF
The corresponding expression for the forward amplitude reads
\BE
  \label{Mgp}
  {\cal M}_{\gamma p}(s,Q^2) = \sum_{\mu,\bar\mu}
  \,\int\limits_0^1 \!\!d\alpha \int d^2\vrT
  \,\Phi^{\!*(\mu,\bar\mu)}_{\Psi}(\alpha,\vrT) \,\sqq(\rT,s)
  \,\Phi^{(\mu,\bar\mu)}_{\gamma}(\alpha,\vrT,Q^2)\,.
\EE
Here the summation runs over spin indexes $\mu$, $\bar\mu$ of the $c$ and
$\bar c$ quarks, $Q^2$ is the photon virtuality, $\Phi_{\gamma}(\alpha,%
\rT,Q^2)$ is the LC distribution function of the photon for a $c\bar c$
fluctuation of separation $\rT$ and relative fraction $\alpha$ of the
photon LC momentum carried by $c$ or $\bar c$. Correspondingly, $\Phi_%
{\Psi}(\alpha,\vrT)$ is the LC wave function of $\Jpsi$ or $\psi'$.
The dipole cross section $\sigma_{q\bar q}(\rT,s)$ mediates the
transition.

This paper is organized as follows. First we review the expressions
of the factorized LC approach to electroproduction of heavy quarkonia
(more details one can find in \cite{HIKT}) and then compare with
available experimental data for $J/\psi$ production. The calculations
which are parameter free demonstrate good agreement with data.

After it we analyse exclusive electroproduction of charmonia off nuclei
$\gamma A\to\Psi X$, where $X=A$ (coherent) or $X=A^*$ (incoherent,
where $A^*$ is an exited state of the $A$-nucleon system). In these
processes new phenomena are to be expected. (i) Color filtering, i.e.
inelastic interactions of the $c\bar c$ pair on its way through the
nucleus is expected to lead to a suppression of $\Psi$ production
relative to $A\sigma_{\gamma p\to\Psi p}$. (ii) Production of a 
$c\bar c$ pair in a nucleus and its absorption are also determined
by the values of the coherence length (lifetime of the $c\bar c$
fluctuation),
\BE
  \label{lc}
  l_c = \frac{2\,\nu}{Q^2+M_{c\bar c}^2} \approx 
        \frac{2\,\nu}{Q^2+M_{\Jpsi}^2}\,,
\EE
where $\nu$ is the energy of the virtual photon in the rest frame of the
nucleus. (iii) Since the dipole cross section $\sqq$ also depends on the
gluon distribution in the target ($p$ of $A$), nuclear shadowing of the
gluon distribution is expected to reduce $\sqq$ in a nuclear reaction
relative to the one on the proton.

The predictions in this paper may be used for planning of future
experiments for electron-nucleus collisions at high energies like in
the eRHIC project. Another possibility to observe photoproduction
off nuclei is heavy ion relativistic collisions (see, for example,
review \cite{BHT}). In the last section we present our results for
$\Jpsi$ production in such processes.


\section{Light-cone dipole approach}

The LC variable describing longitudinal motion of the quarks is
the fraction $\alpha=p_c^+/p_{\gamma}^+$ of the photon LC momentum
$p_{\gamma}^+ = E_{\gamma}+p_{\gamma}$ carried by the quark or
antiquark. $\alpha$ is Lorentz-boost invariant. In the nonrelativistic
approximation (assuming no relative motion of $c$ and $\bar c$)
$\alpha=1/2$ (e.g. \cite{KZ}), otherwise one should integrate
over $\alpha$ (see Eq.~\Ref{Mgp}).

\subsection{Photon wave function}

For transversely ($T$) and longitudinally ($L$) polarized photons  
the perturbative photon-quark distribution function in Eq.~\Ref{Mgp} 
reads \cite{KS,BKS},
\BE
  \label{psi-g}
  \Phi_{T,L}^{(\mu,\bar\mu)}(\alpha,\vrT,Q^2) =
    \frac{\sqrt{N_c\,\alpha_{em}}}{2\,\pi}\,Z_c
    \,\chi_c^{\mu\dagger}\,\widehat O_{T,L}
    \,\widetilde\chi_{\bar c}^{\bar\mu}\,K_0(\epsilon\rT) \ ,
\EE
where $\widetilde\chi_{\bar c} = i\sigma_y\chi^*_{\bar c}$, $\chi$ and
$\bar\chi$ are the spinors of the $c$-quark and antiquark respectively;
$Z_c=2/3$. $K_0(\epsilon\rT)$ is the modified Bessel function with
$\epsilon^2 = \alpha(1-\alpha)Q^2 + m_c^2$. The operators $\widehat%
O_{T,L}$ have the form:
\BA
  \widehat O_{T} &=& m_c \, \vec\sigma\cdot\vec e_\gamma
    + i(1-2\alpha)
      \,(\vec\sigma \cdot \vec n)
      \,(\vec e_\gamma \cdot \vec\nabla_{\rT})
    + (\vec n \times \vec e_\gamma)\cdot\vec\nabla_{\rT}\ ,\\
  \widehat O_{L} &=& 2\,Q\,\alpha(1-\alpha)\,\vec\sigma\cdot\vec n\ ,
\EA
where $\vec n=\vec p/p$ is a unit vector parallel to the photon momentum
and $\vec e$ is the polarization vector of the photon. Effects of the
non-perturbative interaction within the $q\bar q$ fluctuation are 
negligible for the heavy charmed quarks.

\subsection{Charmonium wave function}

The charmonium wave function is well defined in its rest frame where
one can rely on the Schr\"odinger equation. As soon as the rest frame
wave function is known, one may be tempted to apply the Lorentz 
transformation to the $c\bar c$ pair as it would be a classical system
and boost it to the infinite momentum frame. However, quantum effects
are important and in the infinite momentum frame a series of different
Fock states emerges from the Lorentz boost. Therefore the lowest 
$|c\bar c\ra$ component in the infinite momentum frame does not
represent the $|c\bar c\ra$ in the rest frame. We rely on the widely
used procedure \cite{Terent'ev} for the generation of the LC wave
functions of charmonia.

In the rest frame the spatial part of the $c\bar c$ pair wave function 
satisfying the Schr\"odinger equation
\BE
  \label{Schroed}
  \left(-\,\frac{\Delta}{m_c}+V(r)\right)
   \,\Psi_{nlm}(\vr)=E_{nl}
   \,\Psi_{nlm}(\vr)
\EE
is represented in the form
\BE
  \label{wf}
  \Psi(\vr) = \Psi_{nl}(r) \cdot Y_{lm}(\theta,\varphi) \ ,
\EE
where $\vr$ is 3-dimensional $c\bar c$ separation, $\Psi_{nl}(r)$ and
$Y_{lm}(\theta,\varphi)$ are the radial and orbital parts of the wave
function. The following four potentials $V(r)$ have been used
\begin{itemize}
\item ``COR'': Cornell potential \cite{COR},
  \BE
    \label{COR}
    V(r) = -\frac{k}{r} + \frac{r}{a^2}
  \EE
  with $k=0.52$, $a=2.34\GeV^{-1}$ and $m_c=1.84\GeV$.
\item ``BT'': Potential suggested by Buchm\"uller and Tye \cite{BT}
  with $m_c=1.48\GeV$. It has a similar structure as the Cornell
  potential: linear string potential at large separations and 
  Coulomb shape at short distances with some refinements, however.
\item ``LOG'': Logarithmic potential \cite{LOG}
  \BE
    \label{LOG}
    V(r) = -0.6635\GeV + (0.733\GeV) \log(r \cdot 1\GeV)
  \EE
  with $m_c=1.5\GeV$.
\item ``POW'': Power-law potential \cite{POW}
  \BE
    \label{POW}
    V(r) = -8.064\GeV + (6.898\GeV) (r \cdot 1\GeV)^{0.1}
  \EE
  with $m_c=1.8\GeV$.
\end{itemize}
The shapes of the four potentials differ from each other only at large
$r$ ($\geq 1\fm$) and very small $r$ ($\leq 0.05\fm$) separations. Note,
however, that COR and POW use $m_c \approx 1.8\GeV$, while BT and LOG
use $m_c \approx 1.5\GeV$ for the mass of the charmed quark. This
difference will have significant consequences.

For the ground state $1S$ all the potentials provide a very similar behavior
for the radial part $\Psi_{nl}(r)$ at $r>0.3\fm$, while for small $r$ the
predictions differ by up to $30\%$. The peculiar property of the $2S$
state wave function is the node at $r\approx 0.4\fm$ which causes strong
cancellations in the matrix elements Eq.~\Ref{Mgp} and as a result,
a suppression of photoproduction of $\psi'$ relative to $\Jpsi$
\cite{KZ,Benhar}.

The lowest Fock component $|c\bar c\ra$ in the infinite momentum frame
is not related by simple Lorentz boost to the wave function of charmonium
in the rest frame. This makes the problem of building the LC wave function
for the lowest $|c\bar c\ra$ component difficult, no unambiguous solution
is yet known. There are only recipes in the literature, a simple one widely
used \cite{Terent'ev}, is the following. One applies a Fourier transformation
from coordinate to momentum space to the known spatial part of the 
non-relativistic wave function \Ref{wf}, $\Psi(\vr)\Rightarrow\Psi(\vp)$,
which can be written as a function of the effective mass of the $c\bar c$,
$M^2=4(p^2+m_c^2)$, expressed in terms of LC variables
\BE
  M^2(\alpha,\pT) = \frac{\pT^2+m_c^2}{\alpha(1-\alpha)}\ .
\EE 
In order to change integration variable $p_L$ to the LC variable
$\alpha$ one relates them via $M$, namely $p_L=(\alpha-1/2)M(p_T,\alpha)$.
In this way the $c\bar c$ wave function acquires a kinematical factor
\BE
  \label{LC-wf-p}
  \Psi(\vp) \Rightarrow
  \sqrt{2}\,\frac{(p^2+m_c^2)^{3/4}}{(\pT^2+m_c^2)^{1/2}}
  \cdot \Psi(\alpha,\vpT)
  \equiv \Phi_\psi(\alpha,\vpT) \ .
\EE

This procedure was used in \cite{Hoyer} and the result is applied to 
calculation of the amplitudes \Ref{Mgp}. The result was discouraging,
since the $\psi'$ to $\Jpsi$ ratio of the electroproduction cross sections
are far too low in comparison with data. However, the oversimplified
dipole cross section $\sigma_{q\bar q}(\rT)\propto\rT^2$ was used,
and what is even more essential, the important ingredient of Lorentz
transformations, the Melosh spin rotation, was left out.

The 2-dimensional spinors $\chi_c$ and $\chi_{\bar c}$ describing $c$
and $\bar c$ respectively in the infinite momentum frame are known to be
related via the Melosh rotation \cite{Terent'ev,Melosh} to the spinors 
$\bar\chi_c$ and $\bar\chi_{\bar c}$ in the rest frame:
$\bf\overline{\chi}_c        = \widehat R(  \alpha, \vpT)\,\chi_c$ and
$\bf\overline{\chi}_{\bar c} = \widehat R(1-\alpha,-\vpT)\,\chi_{\bar c}$,
where the matrix $R(\alpha,\vpT)$ has the form:
\BE
  \widehat R(\alpha,\vpT) = 
    \frac{  m_c+\alpha\,M - i\,[\vec\sigma \times \vec n]\,\vpT}
    {\sqrt{(m_c+\alpha\,M)^2+\pT^2}} \ .
\label{matrix}
\EE

Since the potentials we use contain no spin-orbit term, the $c\bar c$
pair is in $S$-wave. In this case spatial and spin dependences in the
wave function factorize and we arrive at the following LC wave function
of the $c\bar c$ in the infinite momentum frame
\BE
  \label{LC-wf}
  \Phi^{(\mu,\bar\mu)}_\psi(\alpha,\vpT) =
     U^{(\mu,\bar\mu)}(\alpha,\vpT)\cdot\Phi_\psi(\alpha,\vpT)\ ,
\EE
where
\BE
  U^{(\mu,\bar\mu)}(\alpha,\vpT) = 
    \chi_{c}^{\mu\dagger}\,\widehat R^{\dagger}(\alpha,\vpT)
    \,\vec\sigma\cdot\vec e_\psi\,\sigma_y
    \,\widehat R^*(1-\alpha,-\vpT)
    \,\sigma_y^{-1}\,\widetilde\chi_{\bar c}^{\bar\mu}\,.
\EE

Now we can determine the LC wave function in the mixed longitudinal
momentum - transverse coordinate representation:
\BE
  \label{lc-wf-r}
  \Phi^{(\mu,\bar\mu)}_\psi(\alpha,\vrT) =
    \frac{1}{2\,\pi} 
    \int d^2\vpT\,e^{-i\vpT\vrT}\,
    \Phi^{(\mu,\bar\mu)}_\psi(\alpha,\vpT)\ .
\EE

\subsection{Phenomenological dipole cross section}

The color dipole cross section $\sigma_{q\bar q}(\rT,s)$ is poorly known
from first principles. It is expected to vanish $\propto \rT^2\,\ln\rT$
at small $\rT\!\!\to\!0$ due to color screening \cite{ZKL} and to level
off at large separations. We use a phenomenological form which interpolates
between the two limiting cases of small and large separations. Few
parameterizations are available in the literature, we choose two of them
which are simple, but quite successful in describing data and denote them
by the initials of the authors as ``GBW'' \cite{GBW} and ``KST'' \cite{KST}.

\BA
  \label{GBW}
  \mbox{``GBW'':}~~~~~~~~
  \sigma_{q\bar q}(\rT,x)&=&23.03\left[1-e^{-\rT^2/r_0^2(x)}\right]\mb\ ,\\
  r_0(x) &=& 0.4 \left(\frac{x}{x_0}\right)^{\!\!0.144} \fm\ ,
  \nonumber
\EA
where $x_0=3.04\cdot10^{-4}$. The proton structure function calculated with
this parameterization fits well all available data at small $x$ and in wide
range of $Q^2$ \cite{GBW}. However, it obviously fails describing the hadronic
total cross sections, since it never exceeds the value $23.03\mb$. The
$x$-dependence guarantees Bjorken scaling for DIS at high $Q^2$, however,
Bjorken $x$ is not a well defined quantity in the soft limit. Instead we use
the prescription of \cite{Ryskin}, $x=(M^2_\psi+Q^2)/s$, where $M_\psi$ is
the charmonium mass.

This problem with limited dipole cross section as well as the difficulty with
the definition of $x$ have been fixed in \cite{KST}. The dipole cross section
is treated as a function of the c.m. energy $\sqrt{s}$, rather than $x$,
since $\sqrt{s}$ is more appropriate for hadronic processes. A similarly
simple form for the dipole cross section is used
\BA
  \label{KST}
  \mbox{``KST'':}~~~~~~~~~
  \sigma_{\bar qq}(\rT,s) &=& \sigma_0(s) \left[1 - e^{-\rT^2/r_0^2(s)}
  \right]\ .~~~~~~
\EA
The values and energy dependence of hadronic cross sections are reproduced
with the following expressions
\BA
  \sigma_0(s) &=& 23.6 \left(\frac{s}{s_0}\right)^{\!\!0.08} 
  \left(1+\frac38 \frac{r_0^2(s)}{\left<r^2_{ch}\right>}\right)\mb\ ,\\
  r_0(s)      &=& 0.88 \left(\frac{s}{s_0}\right)^{\!\!-0.14}  \fm\ .
\EA
The energy dependent radius $r_0(s)$ is fitted to data for the proton
structure function $F^p_2(x,Q^2)$, $s_0 = 1000\GeV^2$ and the mean square of
the pion charge radius $\left<r^2_{ch}\right>=0.44\fm^2$. The improvement at
large separations leads to a somewhat worse description of the proton structure
function at large $Q^2$. Apparently, the cross section dependent on energy,
rather than $x$, cannot provide Bjorken scaling. Indeed, parameterization
\Ref{KST} is successful only up to $Q^2\approx 10\GeV^2$. 

In fact, the cases we are interested in, charmonium production and
interaction, are just in between the regions where either of these
parameterization is successful. Therefore, we suppose that the difference
between predictions using Eq.~\Ref{GBW} and \Ref{KST} is a measure of the
theoretical uncertainty which fortunately turns out to be rather small.


\section{Electroproduction off protons}

Having Eq.~\Ref{Mgp} and the expressions from the previous section (LC wave
functions and dipole cross section), we can calculate cross sections for
the virtual photoproduction $\gamma p\to\Psi p$
\BE
  \label{sigma-gp}
  \sigma_{\gamma p\to\Psi p}(s,Q^2) = 
  \frac{\vert\widetilde{\cal M}_T(s,Q^2)\vert^2 + 
  \eps\,\vert\widetilde{\cal M}_L(s,Q^2)\vert^2}{16\,\pi\,B}\ ,
\EE
where $\eps$ is the photon polarization (for H1 data $\la\eps\ra=0.99$); 
$B$ is the slope parameter in reaction $\gamma^* p\to\psi p$. We use the
experimental value \cite{H1-s} $B=4.73\,GeV^{-2}$. $\widetilde{\cal M}_{T,L}$
includes also the correction for the real part of the amplitude:
\BE
  \label{Mgp-t}
  \widetilde{\cal M}_{T,L}(s,Q^2) = {\cal M}_{T,L}(s,Q^2)
   \,\left(1 - i\,\frac{\pi}{2}\,\frac{\partial
   \,\ln{\cal M}_{T,L}(s,Q^2)}{\partial\,\ln s} \right)\ ,
\EE
where we apply the well known derivative analyticity relation between the
real and imaginary parts of the forward elastic amplitude \cite{Bronzan}.
The correction from the real part is not small since the cross section of
charmonium electroproduction is a rather steep function of energy.

\BF
\Ps{0.6}{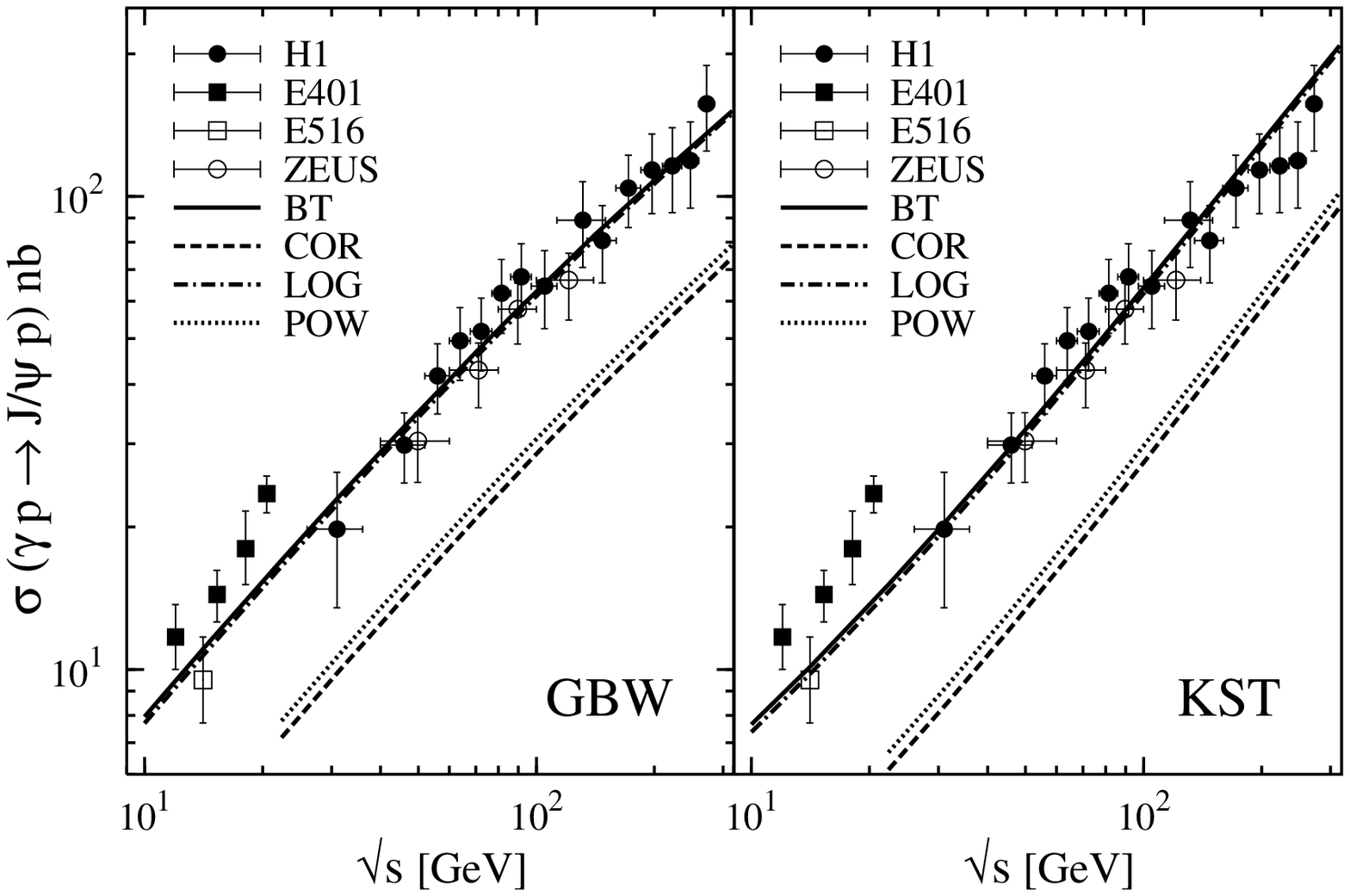}
\caption{
  \label{Fig-s}
  Integrated cross section for elastic photoproduction with real photons
  ($Q^2=0$) calculated with GBW and KST dipole cross sections and for four
  potentials to generate $\Jpsi$ wave functions. Experimental data points
  from the H1~\cite{H1-s}, E401~\cite{E401-s}, E516~\cite{E516-s} and
  ZEUS~\cite{ZEUS-s} experiments.
}
\EF
\BF
\Ps{0.57}{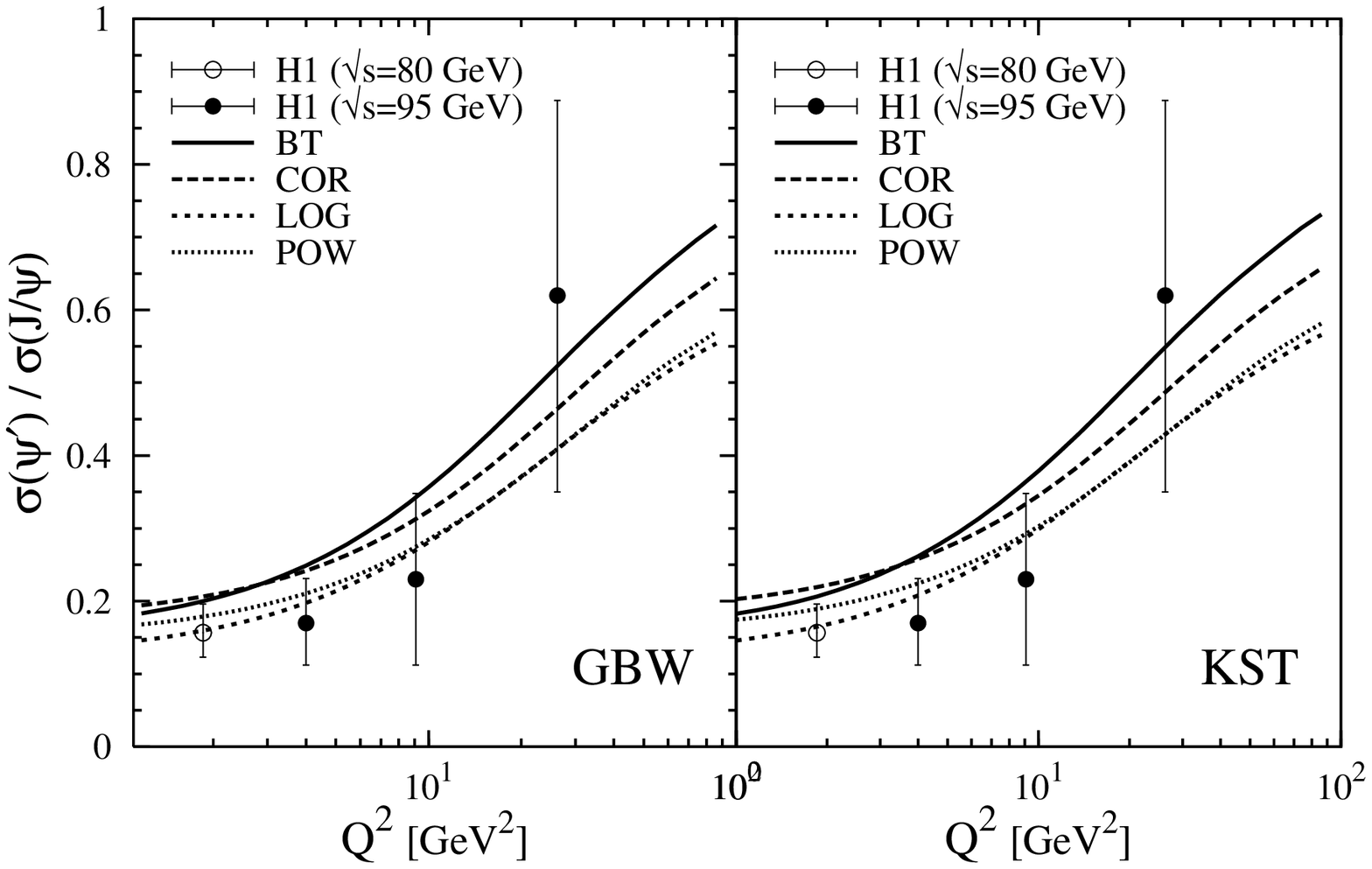}
\caption{
  \label{Fig-R}
  The ratio of $\psi'$ to $J/\psi$ virtual photoproduction cross sections
  as a function of the photon virtuality $Q^2$ at energy $\sqrt{s}=90\GeV$
  for four potentials and with GBW and KST parameterizations for the
  dipole cross section. Experimental data points from the H1 experiment
  \cite{H1-RQ}.
}
\EF

We present here only results for energy dependence of the $\Jpsi$ integrated
cross section (Fig.~\ref{Fig-s}) and for $Q^2$ dependence of the ratio $\psi'$
to $\Jpsi$ photoproduction (Fig.~\ref{Fig-R}). More results are presented in
\cite{HIKT}. All calculations are performed with GBW and KST parameterizations
for the dipole cross section and for wave functions of the $\Jpsi$ calculated
from BT, LOG, COR and POW potentials. One can see that there are no major
differences for the results using the GBW and KST parameterizations. The BT
and LOG potentials describe the data very well, while the potentials COR and
POW underestimate them by a factor of two. The different behavior has been
traced to the following origin: BT and LOG use $m_c \approx 1.5\GeV$, but
COR and POW $m_c \approx 1.8\GeV$. While the bound state wave functions of
$\Jpsi$ are little affected by this difference, the photon wave function
Eq.~\Ref{psi-g} depends sensitively on $m_c$.

It turns out that the effects of spin rotation have a gross impact on
the ratio $R=\sigma(\psi')/\sigma(\Jpsi)$. This effects add 30-40\% to
the $\Jpsi$ electroproduction cross section. But they have a much more
dramatic impact on $\psi'$ increasing the electroproduction cross section
by a factor 2-3. This spin effects explain the large values of the ratio
$R$ observed experimentally. Our results for $R$ are about twice as large
as evaluated in \cite{Suzuki} and even more than in \cite{Hoyer}.


\section{Electroproduction off nuclei}

Exclusive charmonium production off nuclei, $\gamma A \to \Psi X$
is called coherent, when the nucleus remains intact, i.e. $X=A$, or
incoherent, when $X$ is an excited nuclear state which contains
nucleons and nuclear fragments but no other hadrons. The cross
sections depend on the polarization $\eps$ of the virtual photon
(in all figures below we will imply $\eps=1$),
\BE
  \label{sigma-gA}
  \sigma_{\gamma A}(s,Q^2) =
  \sigma_T(s,Q^2) + \eps\,
  \sigma_L(s,Q^2)~,
\EE
where the indexes $T,L$ correspond to transversely or longitudinally 
polarized photons, respectively. At high energies the coherence length
Eq.~\Ref{lc} may substantially exceed the nuclear radius ($l_c\gg R_A$).
In this case the transverse size of the $c\bar c$ wave packet is ``frozen''
by Lorentz time dilation, i.e. it does not fluctuate during propagation
through the nucleus, and the expressions for the incoherent ($inc$) and
coherent ($coh$) cross sections are simple \cite{KZ}:
\BA
  \label{sigma-gA-coh}
  \sigma_{T,L}^{coh}(s,Q^2) &=&
  \int d^2b \left|\widetilde{\cal M}_{T,L}^{coh}(s,Q^2,b)\right|^2\,,\\
  \label{sigma-gA-inc}
  \sigma_{T,L}^{inc}(s,Q^2) &=&
  \int d^2b \left|\widetilde{\cal M}_{T,L}^{inc}(s,Q^2,b)\right|^2\,
  \frac{T_A(b)}{16\pi B(s)}\,,
\EA
where $T_A(b)=\int_{-\infty}^{\infty}dz\,\rho_A(b,z)$ is the nuclear
thickness function given by the integral of the nuclear density along
the trajectory at a given impact parameter $b$ and expressions for
$\widetilde{\cal M}(s,Q^2,b)$ are given by Eqs.~\Ref{Mgp-t} and
\Ref{Mgp} with replacement
\BE
  \label{frozen}
  \sqq(r_T,s) \Rightarrow \left\{
  \begin{array}{rl}
           1 - \exp\left[-\sqq(r_T,s)\,T_A(b)/2\right],& coh.\\[2mm]
    \sqq(r_T,s)\exp\left[-\sqq(r_T,s)\,T_A(b)/2\right],& inc.
  \end{array}
  \right.
\EE

But for charmonium production off nuclei the ``frozen'' approximation
is not enough and additional phenomena should be taken into account:
effects of finite coherence length and gluon shadowing.

\subsection{Finite coherence length}

The ``frozen'' approximation \Ref{frozen} is valid only for $l_c\gg R_A$
and can be used only at $\sqrt{s} > 20\div30\,\GeV$. The low-energy part
should be corrected for the effects related to the finiteness of $l_c$.
A strictly quantum-mechanical treatment of a fluctuating $q\bar q$ pair
propagating through an absorptive medium based on the LC Green function
approach has been suggested recently in \cite{KNST}. However, an analytical
solution for the LC Green function is known only for the simplest form of
the dipole cross section $\sqq(r_T) \propto r_T^2$. With a realistic form
of $\sqq(r_T)$ it is possible only to solve this problem numerically,
what is still a challenge. Here we use the approximation suggested in
\cite{HKZ} to evaluate the corrections arising from the finiteness of
$l_c$ by multiplying the cross sections for coherent and incoherent
production evaluated for $l_c\to\infty$ by a kind of formfactor
$F^{coh}$ and $F^{inc}$ respectively:
\BE
  \sigma_{\gamma A}(s,Q^2) \Rightarrow
  \sigma_{\gamma A}(s,Q^2) \cdot F\left(s,l_c(s,Q^2)\right)\,,
\EE
where
\BA
  \label{Fcoh}
  F^{coh}(s,l_c) &=& \int\!d^2 b
    \left|
    \,\int\limits_{-\infty}^\infty\!dz\,\rho_A(b,z)
    \,F_1(s,b,z)\,e^{iz/l_c}
    \right|^2 / \,(...)|_{l_c = \infty} \,,\\
  \label{Finc}
  F^{inc}(s,l_c) &=& \int\!d^2 b
    \,\int\limits_{-\infty}^\infty\!dz\,\rho_A(b,z)
    \,\left|F_1(s,b,z)-F_2(s,b,z,l_c)\right|^2 / \,
    (...)|_{l_c = \infty} \,,\\
  F_1(s,b,z)     &=&
    \,\exp\left(-\frac12\,\sigma_{\Psi N}(s)
    \!\int\limits_z^\infty\!dz'\,\rho_A(b,z')\right)\,,\\
  F_2(s,b,z,l_c) &=& \frac12\,\sigma_{\Psi N}(s)
    \!\int\limits_{-\infty}^z\!dz'
    \,\rho_A(b,z')\,F_1(b,z')\,e^{-i(z-z')/l_c}\,.
\EA
For the charmonium nucleon total cross section $\sigma_{\Psi N}(s)$ we use
our previous results \cite{HIKT}.

\subsection{Gluon shadowing}

The gluon density in nuclei at small Bjorken $x$ is expected to be
suppressed compared to a free nucleon due to interferences. This
phenomenon, called gluon shadowing, renormalizes the dipole cross section,
\BE
  \label{gluon}
  \sqq(r_T,x) \Rightarrow \sqq(r_T,x)\,R_G(x,Q^2,b)\,.
\EE
where the factor $R_G(x,Q^2,b)$ is the ratio of the gluon density at $x$
and $Q^2$ in a nucleon of a nucleus to the gluon density in a free nucleon.
No data are available so far which could provide information about gluon
shadowing. Currently it can be evaluated only theoretically. To calculate
function $R_G$ we use the approach developed in \cite{KST} and applied for
charmonia production in \cite{HIKT-A}.

\subsection{Numerical results}

Combining the results above (i.e. including finite coherence length and
gluon shadowing) we obtain the results for charmonia ($\Jpsi$ and $\psi'$)
electroproduction off nuclei. As it is common practice we express nuclear
cross sections in the form of the ratio
\BE
  \label{R-def}
  R_\Psi(s,Q^2) =
  \frac{\sigma_{\gamma A}(s,Q^2)}{A\,\sigma_{\gamma p}(s,Q^2)}~,
\EE
where the numerator stands for the expression Eq.~\Ref{sigma-gA}
(with Eq.~\Ref{sigma-gA-coh} for coherent and Eq.~\Ref{sigma-gA-inc}
for incoherent scattering) and the denominator is given by Eq.~\Ref{sigma-gp}.
We present here only $s$ dependences (plots for $Q^2$ dependences and
momentum transfer $\vec k_T$ distributions are given in \cite{HIKT-A})
for coherent (Fig.~\ref{s-coh-full}) and incoherent (Fig.~\ref{s-inc-full})
production of charmonia with the GBW \cite{GBW} parameterization for the
dipole cross section. KST \cite{KST} parameterization gives quite close
results (they differ at most 10\% at high energies). It is not a surprise
that the ratios for coherent production exceed one: in the absence of
$c\bar c$ attenuation the forward coherent production would be proportional
to $A^2$, while integrated over momentum transfer it behaves as $A^{4/3}$.
It is a result of our definition Eq.~\Ref{R-def} that $R^{coh}_{\Psi}$
exceeds one.
\BF
\PS{0.6}{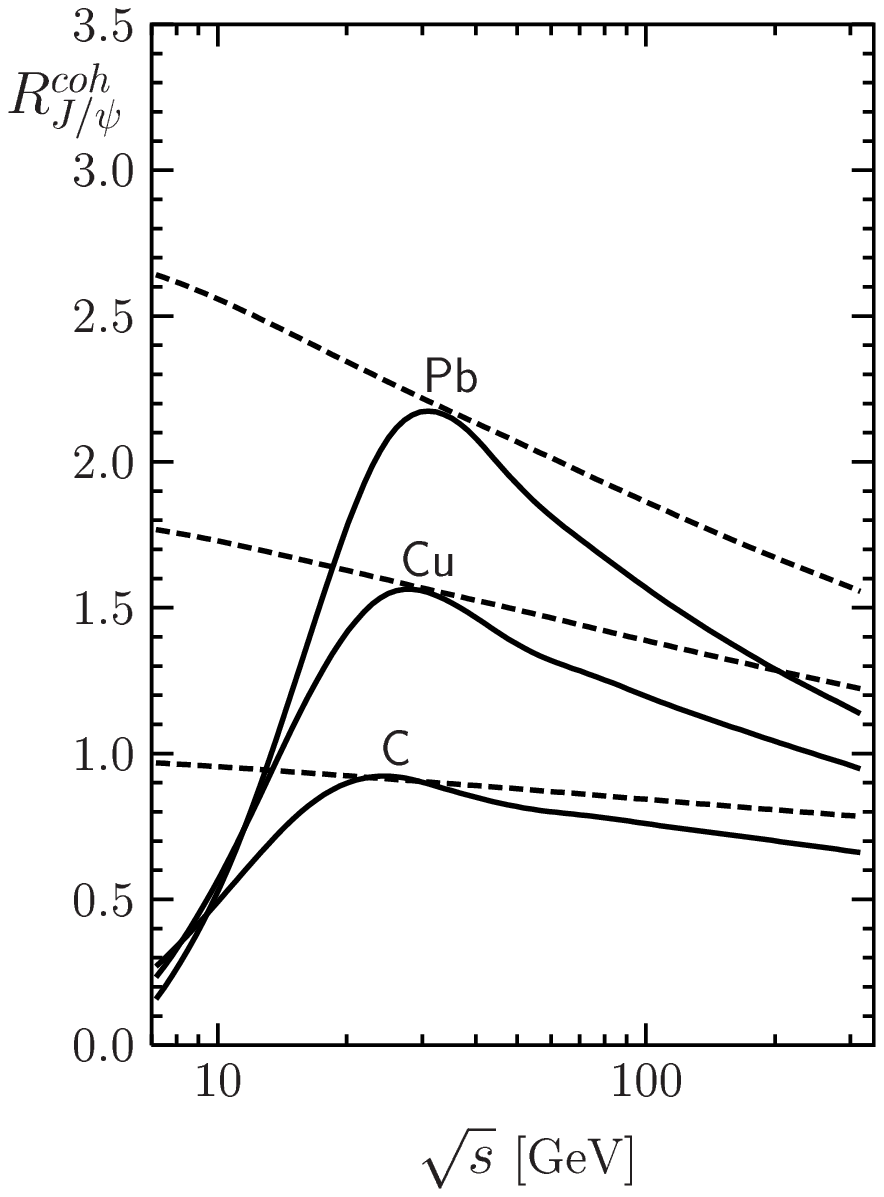}{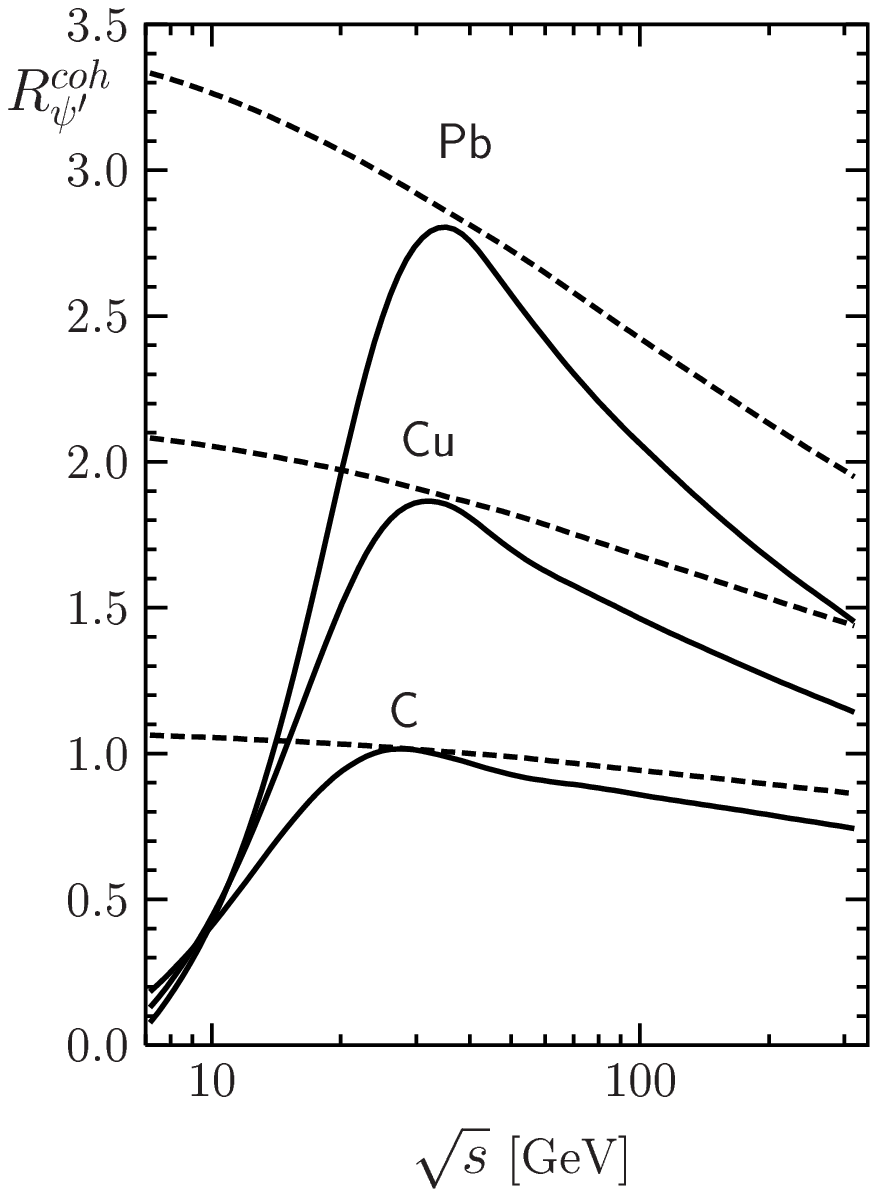}
\caption{
  \label{s-coh-full}
  Ratios $R^{coh}_{\Jpsi}$ and $R^{coh}_{\psi'}$ for coherent production
  on carbon, copper and lead as a function of $\sqrt s$ and at $Q^2=0$
  calculated with GBW parameterization of $\sqq$. Solid curves display
  the modifications due to the finite coherence length $l_c$ and gluon
  shadowing corrections while the dashed lines are without (``frozen''
  approximation).
}
\EF
\BF
\PS{0.6}{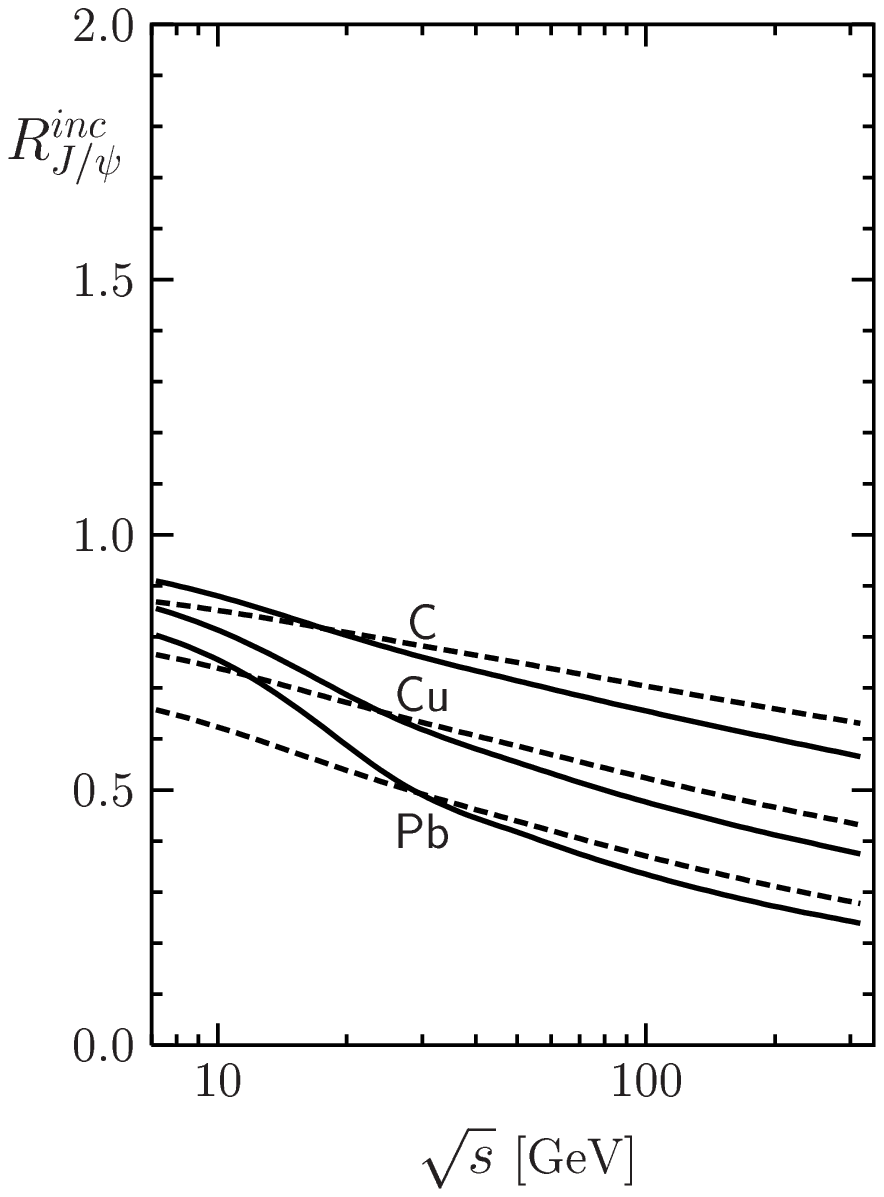}{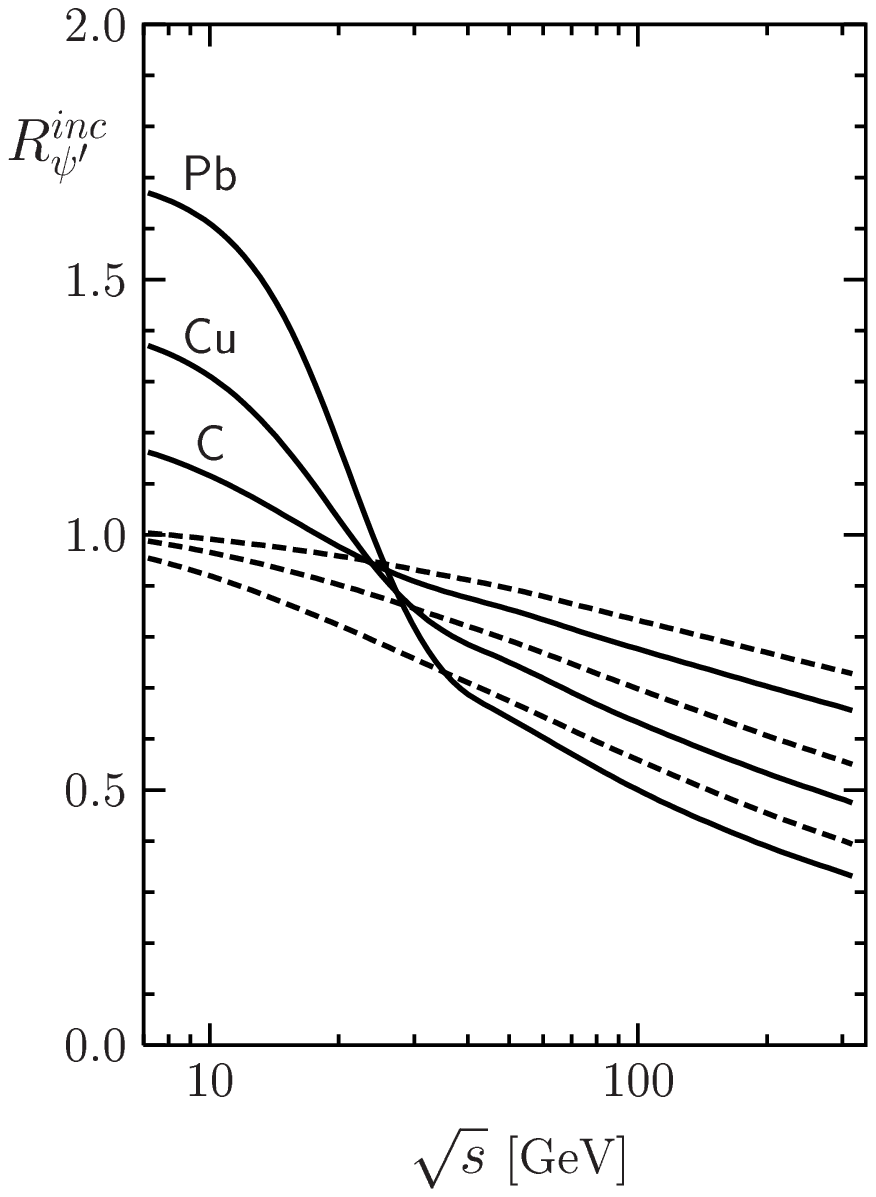}
\caption{
  \label{s-inc-full}
  Ratios $R^{inc}_\Psi$ for $\Jpsi$ and $\psi'$ incoherent production
  on nuclei as function of $\sqrt s$ and at $Q^2=0$. The meaning of the
  different lines is the same as in Fig.~\ref{s-coh-full}.
}
\EF

We see that in comparison with the ``frozen'' approximation, $l_c$
corrections noticeable change ratios at low energies (especially for
$\psi'$). For coherent production in the limit of low energies $l_c \to 0$
the strongly oscillating exponential phase factor in (\ref{Fcoh}) makes
the integral very small and thus $F^{coh}\approx0$. Then the cross section
rises with $l_c$ unless it saturates at $l_c \gg R_A$ when the phase factor
becomes constant. Apparently, this transition region is shifted to higher
energies for larger nuclear radius. For incoherent production the observed
nontrivial energy dependence is easy to interpret. At low energies $l_c \ll R_A$
and the photon propagates without any attenuation inside the nucleus where it
develops for a short time $t_c$ a $c\bar c$ fluctuation which momentarily
interacts to get on mass shell. The produced $c\bar c$ pair attenuates along
the path whose length is a half of the nuclear thickness on the average. On
the other hand, at high energies when $l_c\gg R_A$ the $c\bar c$ fluctuation
develops long before its interaction with the nucleus. As a result, it
propagates through the whole nucleus and the mean path is twice as long
as at low energies. This is why the nuclear transparency drops when going
from the regimes of short to long $l_c$. 

At high energies gluon shadowing becomes important. We see that the onset
of gluon shadowing happens at a c.m. energy of few tens of GeV. Remarkably,
the onset of shadowing is delayed with rising nuclear radius. Nuclear
suppression of $\Jpsi$ production becomes stronger with energy. This is
an obvious consequence of the energy dependence of $\sqq(r_T,s)$, which
rises with energy. For $\psi'$ the suppression is rather similar as for
$\Jpsi$. In particular we do not see any considerable nuclear enhancement
of $\psi'$ which has been found earlier \cite{KZ,KNNZ}, where the
oversimplified form of the dipole cross section, $\sqq(r_T)\propto r_T^2$
and the oscillator form of the wave function had been used.


\section{Heavy ion peripheral collisions}

The large charge $Z$ of heavy nuclei gives rise to strong electromagnetic
fields: the photon field of one nucleus can produce a photo-nuclear interaction
in the other. The cross section of the charmonia photoproduction by the induced
photons reads

\BE
  k\frac{d\sigma}{dk} = \int\!\!d^2b \int\!\!d^2b'\,
  n(k,\vec b'-\vec b)\, \sigma_A(b,s)\,,
\EE
where $k$ is the photon momentum. Photon flux induced by projectile
nucleus with Lorenz factor $\gamma$ is
\BE
  n(k,\vec b) = \frac{\aem Z^2 k^2}{\pi^2{\gamma}^2}
  K_1^2\left(\frac{bk}{\gamma}\right)\,.
\EE
Cross sections $\sigma_A(b,s)$ for coherent and incoherent production are
\BA
  \sigma_A^{coh}(s,b) &=& \left|\widetilde{\cal M}_A^{coh}(s,b)\right|^2\,,\\
  \sigma_A^{inc}(s,b) &=& \left|\widetilde{\cal M}_A^{inc}(s,b)\right|^2
                          \,\frac{T(b)}{16 \pi B(s)}\,,
\EA
where expressions $\widetilde{\cal M}(s,b)$ correspond to $\widetilde{\cal M}$
in Eqs.~(\ref{sigma-gA-coh}, \ref{sigma-gA-inc}) at $Q^2=0$. Our predictions
for coherent $\Jpsi$ production at RHIC and LHC energies are presented on
Fig.~\ref{PC}.
\BF
\PS{0.6}{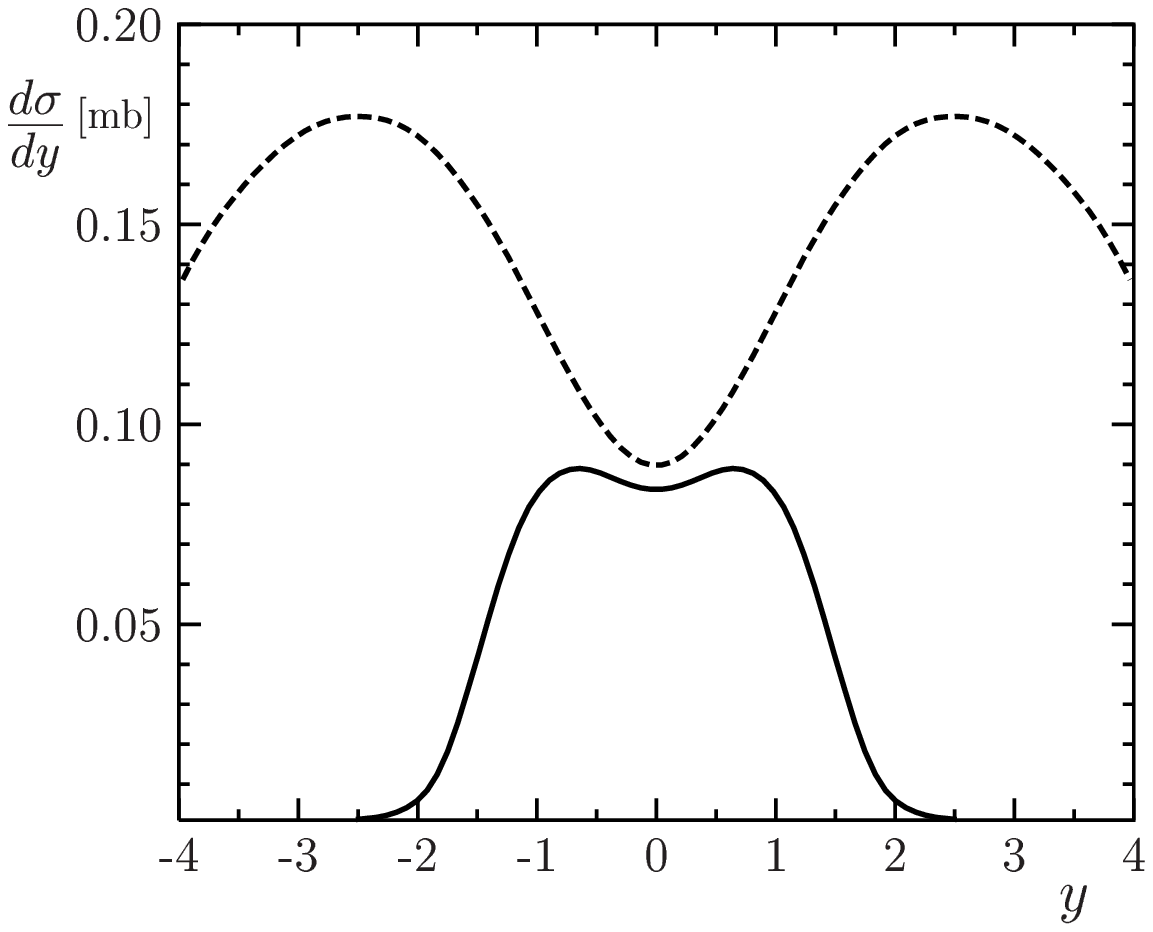}{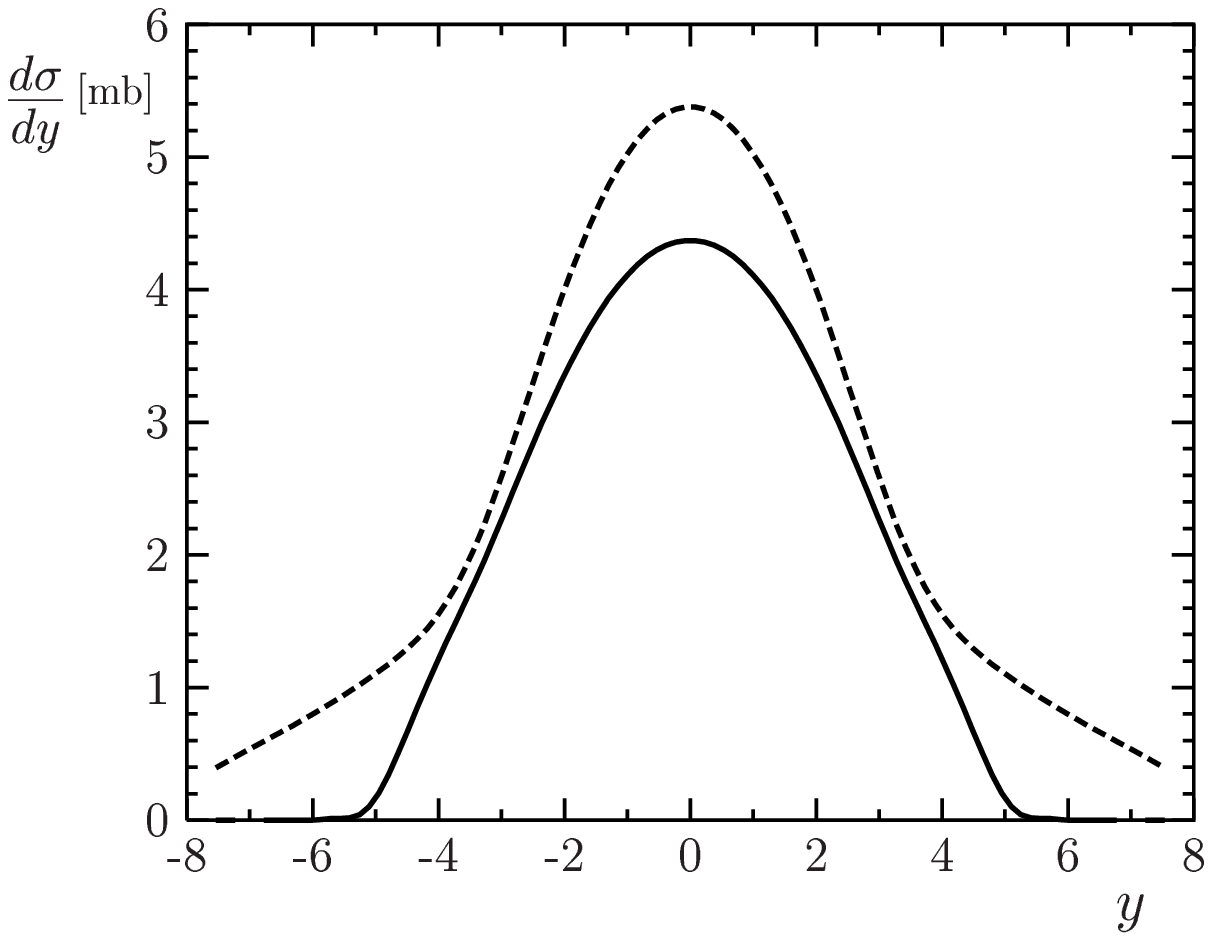}
\caption{
  \label{PC}
  Rapidity distribution for coherent $\Jpsi$ production in heavy ion peripheral
  gold-gold collisions at RHIC (left) and lead-lead collisions at LHC (right)
  calculated with the GBW parameterization of $\sqq$. Solid curves display the
  modifications due to gluon shadowing and finite coherence length $l_c$
  while the dashed lines are without (``frozen'' approximation).
}
\EF
We see that $l_c$ corrections modify the distribution at the edges (positive and
negative $y$) while suppression around $y=0$ is provided mostly by gluon shadowing
(especially for LHC, where energies should be much higher).

\section{Conclusion}

In this paper we use the LC dipole approach for description of charmonium
electroproduction off protons and nuclei. We have no free parameters and
our calculations are in good agreement with existing experimental data.
Our predictions can be tested in future experiments at high energies
with electron-nuclear colliders (eRHIC) and in peripheral heavy ion
collisions (RHIC and LHC).

\noindent {\bf Acknowledgment}: The authors gratefully acknowledge the
partial support by a grant from the Gesellschaft f\"ur Schwerionenforschung
Darmstadt (grant no.~GSI-OR-SCH) and by the Federal Ministry BMBF (grant
no.~06~HD~954).


\end{document}